\documentclass[aps,prb,twocolumn,superscriptaddress,showpacs] {revtex4-1} 

\usepackage{natbib}
\usepackage{graphicx}
\usepackage{dcolumn}
\usepackage{textcomp}
\usepackage{bm}
\usepackage{color}
\begin{document}
\title{Quantum oscillations of the superconductor LaRu$_2$P$_2$ : comparable mass enhancement $\lambda \approx 1$ in Ru and Fe phosphides}

\author{Philip J.W. Moll} 
\author{Jakob Kanter}
\affiliation{Laboratory for Solid State Physics, ETH Zurich, Switzerland} 
\author{Ross D. McDonald}
\author{Fedor Balakirev}
\affiliation{National High Magnetic Field Laboratory, Los Alamos National Laboratory, MS-E536, Los Alamos, New Mexico 87545, USA} 
\author{Peter Blaha}
\author{Karlheinz Schwarz}
\affiliation{Institute of Materials Chemistry, Vienna University of Technology, Austria}
\author{Zbigniew Bukowski}
\author{Nikolai D. Zhigadlo}
\author{Sergiy Katrych}
\author{Kurt Mattenberger} 
\author{Janusz Karpinski} 
\author{Bertram Batlogg}
\affiliation{Laboratory for Solid State Physics, ETH Zurich, Switzerland}

\begin{abstract}

We have studied the angular dependent de Haas-van Alphen oscillations of LaRu$_2$P$_2$ using magnetic torque in pulsed magnetic fields up to 60T. The observed oscillation frequencies are in excellent agreement with the geometry of the calculated Fermi surface. The temperature dependence of the oscillation amplitudes reveals effective masses m*($\alpha$)=0.71 and m*($\beta$)=0.99 m$_e$, which are enhanced over the calculated band mass by $\lambda^{cyc}$ of 0.8. We find a similar enhancement $\lambda^{\gamma} \approx 1$ in comparing the measured electronic specific heat ($\gamma = 11.5$ mJ/mol K$^2$) with the total DOS from band structure calculations. Remarkably, very similar mass enhancements have been reported in other pnictides LaFe$_2$P$_2$, LaFePO ($T_c \approx 4K$), and LaRuPO, independent of whether they are superconducting or not. This is contrary to the common perceptions that the normal state quasi-particle renormalizations reflect the strength of the superconducting paring mechanism and leads to new questions about pairing in isostructural and isoelectronic Ru- and Fe-pnictide superconductors.

\end{abstract}

\pacs{71.18.+y,74.70.Xa}

\maketitle

Electronic correlations and the associated electron mass enhancements are a central aspect in the discussion of superconductivity in transition metal compounds\cite{Basov2009}. Correlations are of particular importance in unconventional superconductors such as the cuprates and pnictides, which are not dominated by electron-phonon coupling \cite{Igor}. As superconductivity in the pnictides can be tuned by changing various parameters such as carrier doping \cite{Kamihara,Zhao2009}, chemical \cite{Wang,Palenzona2009} and external pressure \cite{Okada,Lorenz2009}, valuable insights on the pairing mechanism may be gained by direct comparison of closely related systems. A particularly intriguing situation presents itself when two seemingly similar pairs of isostructural and isoelectronic compounds and their superconductivity are compared: LaRu$_2$P$_2$ –-- LaFe$_2$P$_2$ and LaRuPO --– LaFePO: While in the former, LaRu$_2$P$_2$ is superconducting ({\it T}$_c$ = 4K)  and LaFe$_2$P$_2$ is not, in the latter the roles of Fe(3d) and Ru(4d) are interchanged and LaFePO is superconducting ({\it T}$_c$ = 4K) while LaRuPO is not. To approach this puzzle, we study in detail the electronic structure near E$_F$ and the enhancement of the quasiparticle mass over thecalculated bare band mass.\\
Experimentally, we have measured quantum oscillations in LaRu$_2$P$_2$ single crystals by means of torque magnetometry in high magnetic fields up to 60 T. Quantum oscillations are a precise bulk tool to identify the Fermi surface geometry and to measure band specific effective masses on the extremal orbits \cite{Shoenberg}. Furthermore, we have measured the heat capacity to determine the electronic density of states (DOS) at E$_F$. Theoretically, we have calculated the electronic band structure of LaRu$_2$P$_2$ and LaFe$_2$P$_2$ using the density functional theory (DFT) code WIEN2k \cite{Blaha,Schwarz2010} to identify similarities and differences.
Importantly, the comparison of quantum oscillations, specific heat data and band calculations allows to estimate the individual band contributions to the many-body mass enhancement.
\begin{figure}[!h]
\centering
\includegraphics[width =0.45 \textwidth]{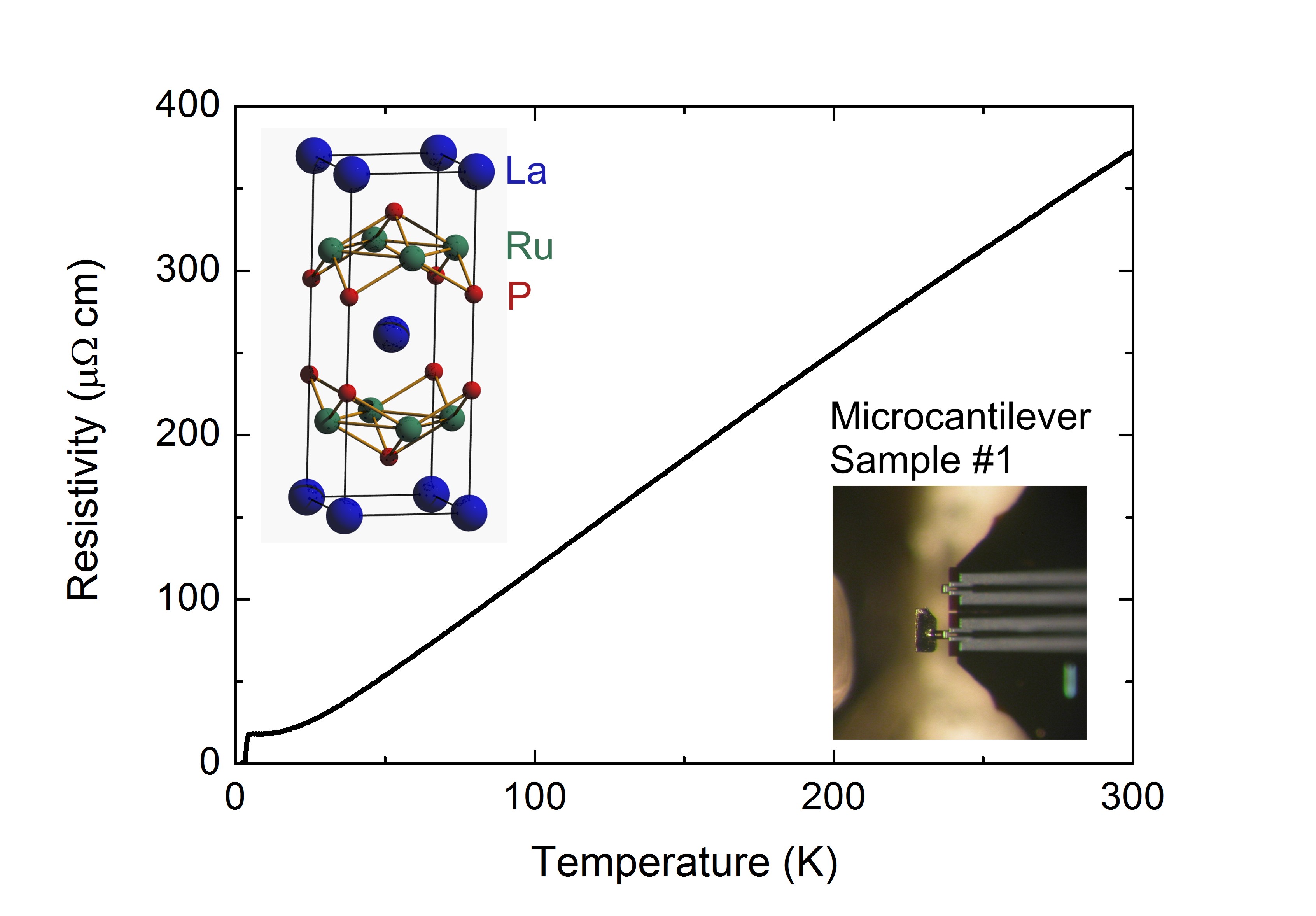}
\caption{In-plane resistivity of a typical LaRu$_2$P$_2$ single crystal as a function of temperature. The four-point resistivity begins to saturate below 20 K, giving a residual resistivity ratio of 21. Left inset: LaRu$_2$P$_2$ unit cell (ThCr$_2$Si$_2$-type). Right inset: Sample 1 mounted on a SEIKO cantilever. }
\label{fig:Resistance}
\end{figure}

\begin{figure}[!h]
\centering
\includegraphics[width =0.51 \textwidth]{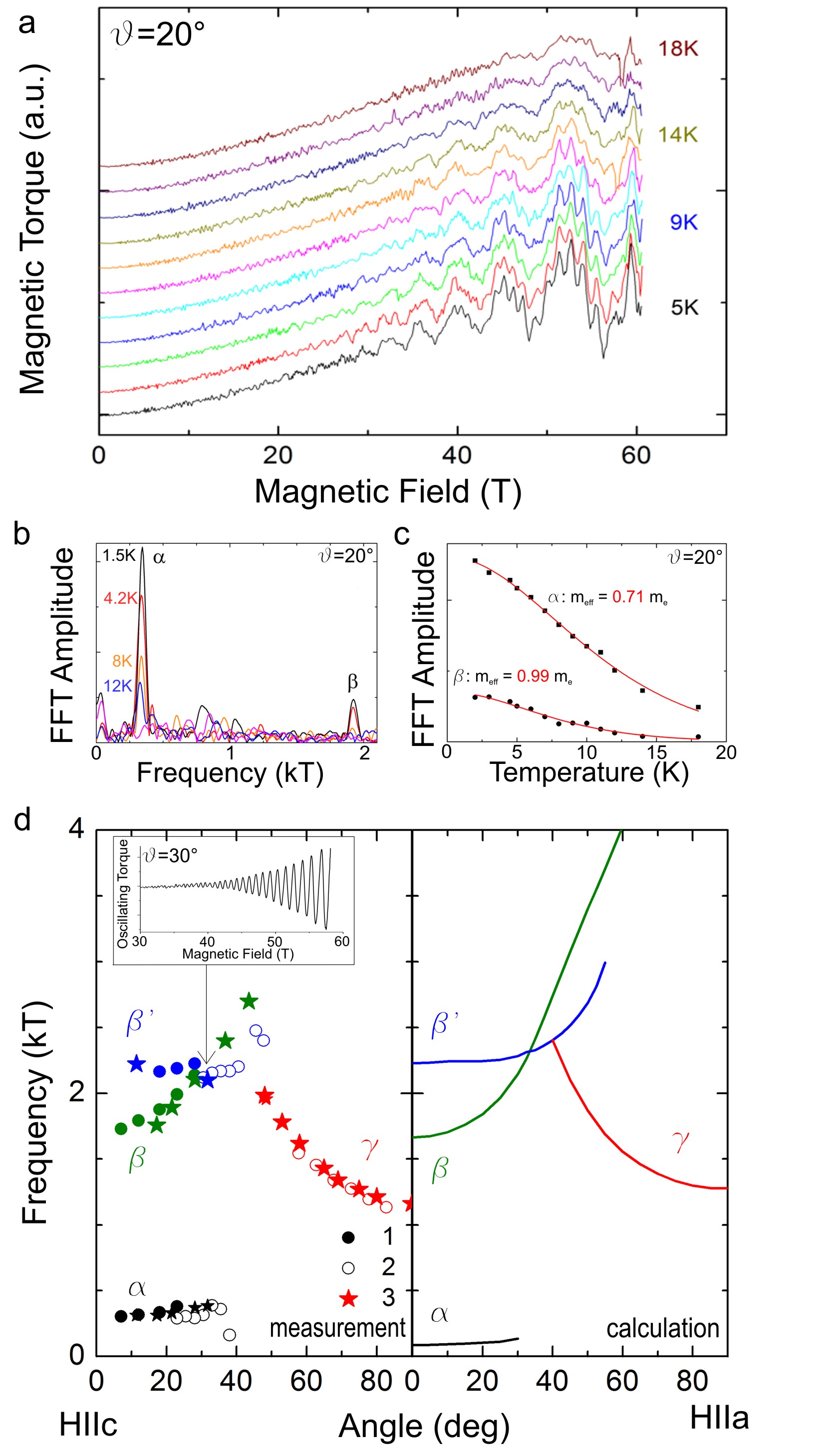}
\caption{a) Raw torque signal measured at various temperatures between 18 K and 5K at an angle $\theta$ = 20\textdegree  between field and {\it c}-axis (curves are offset for clarity). The quantum oscillations are clearly visible at the high temperature of 18 K, indicating the light effective masses on these orbits. b) Frequency spectrum of this data. c) Temperature dependence of the oscillatory torque amplitudes ($\theta$ = 20\textdegree), originating from the $\alpha$ (Donut Hole) and $\beta$ (Cylinder) orbits. The line shows a fit to the Lifshitz-Kosevich formula. d) Measured and calculated angular dependence of the amplitudes. The good agreement allows for the identification of orbits on the FS.}
\label{fig:Amplitudes}
\end{figure}

\begin{figure}[h]
\centering
\includegraphics[width =0.48 \textwidth]{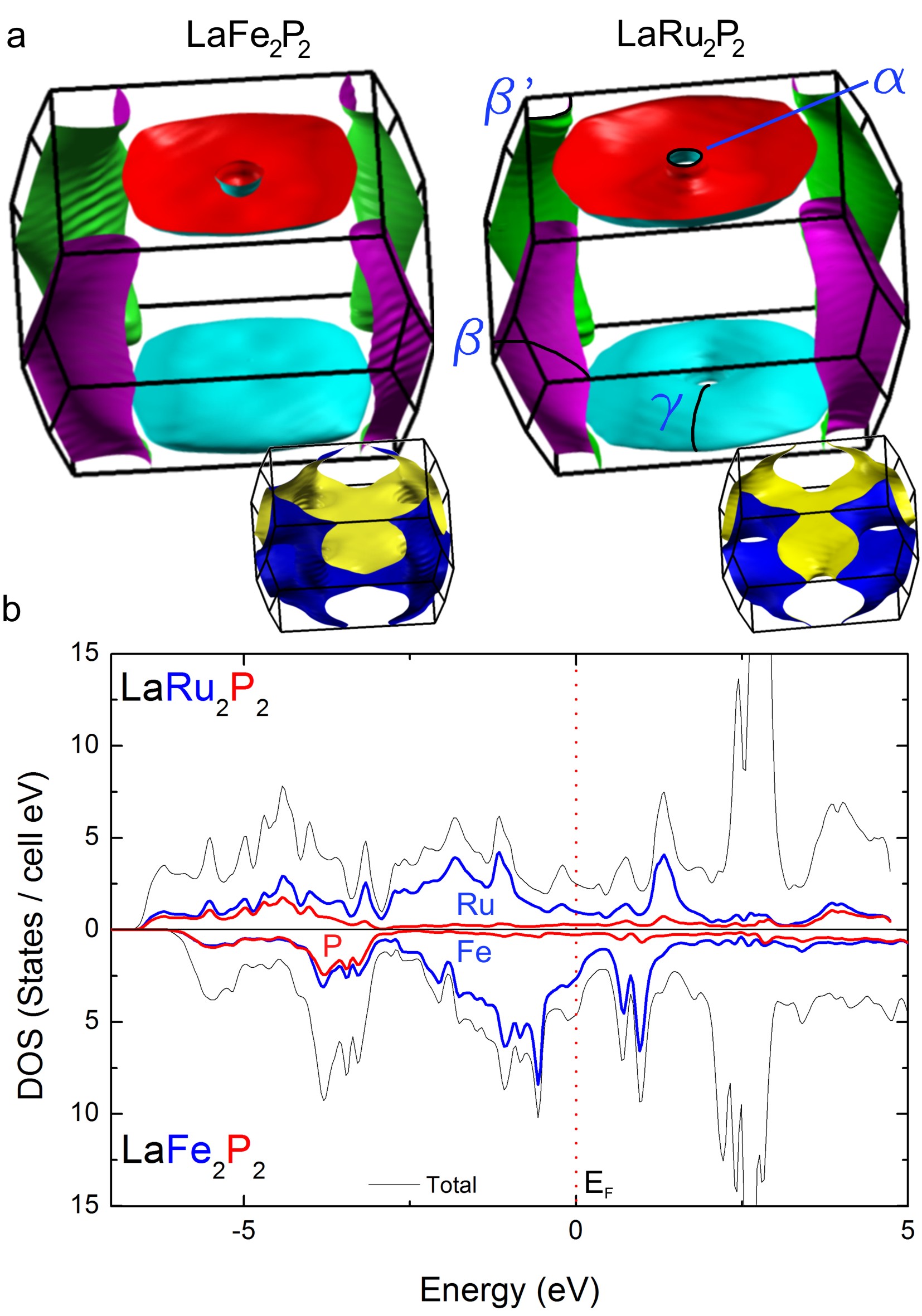}
\caption{a) Fermi surfaces of LaFe$_2$P$_2$ and LaRu$_2$P$_2$ calculated by DFT. The orbits observed in the dHvA experiments in LaRu$_2$P$_2$ have been labeled. b) Density of states comparison between the Fe and Ru compound. The more localized Fe 3d states cause in LaFe$_2$P$_2$ a smaller bandwidth and a larger DOS at the Fermi level.}
\label{fig:Calculation}
\end{figure}

Single crystals of LaRu$_2$P$_2$ were grown by the Sn flux method using LaP$_2$ and Ru powders as starting materials, similar to other '122' pnictides\cite{Bukowski2009}. The components (1:2:2:40 La:Ru:P:Sn ratio) were loaded into an alumina crucible and sealed in an evacuated silica tube. The ampoule was kept at 1100 \textdegree C for 12 hours, slowly cooled to 750 \textdegree C at 1.5-2 \textdegree C/h followed by fast cooling to room temperature. The Sn matrix was dissolved in hydrochloric acid. Powder X-ray diffraction analysis performed on crushed crystals confirmed the ThCr$_2$Si$_2$-type structure with the lattice parameters {\it a} = 4.025 \AA$ $ and {\it c} = 10.662 \AA. We have studied three individual crystals from the same growth batch. The large crystals were cut into suitable pieces of about 200 x 100 x 40 $\mu$m size. Crystals 1 and 2 were studied as grown, while crystal 3 was sealed in a quartz ampoule and annealed in vacuum at 800\textdegree C for 10 days to further improve the crystal quality. All three crystals gave quantum oscillations of the same frequency and of comparable amplitude. The in-plane residual resistivity ratio $\rho(300K)$/$\rho(5K)$ was typically about 20 (Fig.\ref{fig:Resistance}). Polycrystalline samples of LaFe$_2$P$_2$ were prepared under high pressure (30 kbar) at 1400 \textdegree C, and phase purity was checked by X-Ray. The high field measurements up to 60T with a pulse duration of 200ms were performed at the National High Magnetic Field Laboratory (NHMFL/LANL). The sample was glued to a commercial piezoresistive microcantilever (SEIKO PRC-120, lower left inset Fig. \ref{fig:Resistance}). These devices detect sensitively oscillations in the magnetic torque $\tau \propto M \times H$ \cite{Osada2002}. The cantilever resistance was measured in a balanced Wheatstone configuration at 297.5 kHz.\\
Figure \ref{fig:Amplitudes}a shows the raw torque signal with tilt angle of $\theta$ = 20\textdegree $ $ between the field and the crystallographic {\it c}-axis (perpendicular to the Ru-P Layers) at temperatures between 5 and 18 K. The weakly paramagnetic background is subtracted using a smooth polynomial of 3rd order to obtain the oscillatory part of the torque and to suppress 1/f noise in the Fourier spectrum (Fig. \ref{fig:Amplitudes}b). At this angle, we find two main frequency components, $\alpha$ (349 T) and $\beta$ (1921 T). The reduction of the oscillation amplitudes A with increasing temperature due to the broadening of the Fermi-Dirac distribution is well described by the Lifshitz-Kosevich formalism, $A/A_0 = X/\sinh(X)$ with $X= 2 \pi^2 k_B T / \hbar \omega_c$ (Fig. \ref{fig:Amplitudes}c). From this fit, we extract effective masses m* for the two orbits as $m_\alpha^{*} = 0.71$ $m_e$ and $m_\beta^*=0.99$ $m_e$. While these appear to be light masses, these values have to be compared later with calculated bare band masses.\\
Now we turn to the FS geometry. The frequencies F are directly connected to the k-space area $S_k$ encompassed by an extremal cyclotron orbit on the FS perpendicular to the applied field via the Onsager relation $F = 2 \pi e S_k / \hbar$. Thus studying the angular dependence of the frequencies allows to determine the bulk FS tomographically. The rotation axis was aligned with the crystal facet perpendicular to the {\it a}- or equivalent {\it b}-axis, to select a magnetic field rotation from the {\it c}-axis ($\theta$ = 0\textdegree) to the {\it a}-axis ($\theta$ = 90\textdegree). The frequencies of all three studied samples agree very well and are given in Fig. \ref{fig:Amplitudes}d. At small angles, the spectrum is given by the low lying $\alpha$ and the higher $\beta$ and $\beta '$ frequencies. At higher angles, between 40\textdegree and 50\textdegree, the spectrum changes and only one frequency $\gamma$ is present.\\
To interpret this data, we have performed electronic band structure calculations using WIEN2k in the PBE-GGA\cite{Perdew}. The LaRu$_2$P$_2$ structure was calculated from experimental X-ray data \cite{Jeitschko} without lattice relaxation as no notable forces on the atoms resulted from the use of these coordinates. The calculated Fermi surface of LaRu$_2$P$_2$ is given in Fig. \ref{fig:Calculation}a. Three bands cross E$_F$ and produce three FS sheets: 1) A donut with a small hole centered around the M point (outside: turquoise, inside: red), 2) a warped electron cylinder around the zone corners (outside: green, inside: purple) and 3) a strongly folded three-dimensional open FS (for clarity moved to the side, blue and yellow).\\
From these FS geometries we calculated the extremal cross-sections for the various angles and express them as quantum oscillation frequencies shown in Fig. \ref{fig:Amplitudes}d. The good agreement between calculated and measured frequencies allows us to identify them with orbits on the FS. The $\beta$ and $\beta '$ orbits correspond to the neck and belly of the cylinder around X. These two frequencies get closer at higher angles, and finally cross at an Yamaji angle ref \cite{Yamaji1989,Shoenberg} of $\theta$ = 30\textdegree (marked in Fig.\ref{fig:Amplitudes}d). This is very close to the calculated value of 33\textdegree and the excellent agreement is a strong indication that the calculation accurately reproduces the warping of the cylinder. The cylinder corresponds to the hybridized Ru-4d,P-2p in-plane bonds, and is a common feature in tetragonal '122' pnictide superconductors (e.g. \cite{Analytis,Terashima,Coldea}). The $\gamma$ branch between 50\textdegree $ $ and 90\textdegree $ $ agrees within a few percent with the cross-sectional orbit of the donut FS. The $\alpha$ orbit corresponds to the small donut hole, which is about 3.5 times larger in area than calculated. The hole diameter is very sensitive to shifts of the Fermi energy, and can be reconciled with the data by a small energy shift of about 40 meV.\\
The measured mass m$_{\beta}$ of 0.99 m$_e$ is light compared to the masses of the electron cylinders centered at X in other isotructural '122' phosphorus compounds:  $m^*/m_e = 1.5 $ (BaNi$_2$P$_2$ \cite{Terashima}), 2.05 (CaFe$_2$P$_2$ in CT-Phase \cite{Coldea}) and 1.6-2.1 (SrFe$_2$P$_2$ \cite{Analytis}). Interestingly, BaFe$_2$As$_2$, the parent compound of the (Ba,K)Fe$_2$As$_2$ high temperature superconductor, shows similar light masses around 0.9 m$_e$ \cite{Analytis2} in the orthorhombic low temperature phase \cite{Huang}.\\
To better understand the inherent difference between the Fe and Ru analogues, we have contrasted in Fig. \ref{fig:Calculation} the calculated electronic structure of LaFe$_2$P$_2$ and LaRu$_2$P$_2$. The two FS are indeed very similar, where the former has a slightly thicker cylinder and the donut hole becomes a topologically disconnected sphere enclosed in a solid pillow. Differences in the electronic structure, however, become evident when the over all band structure is considered. In Fig. \ref{fig:Calculation}b we contrast the DOS of the two materials. The replacement of Fe-3d by less localized Ru-4d orbitals increases the bandwith and thus lowers the overall DOS. At the Fermi level, we calculate 4.60 states/eV in LaFe$_2$P$_2$ and 2.46 states/eV in LaRu$_2$P$_2$. These values correspond to a linear electronic heat coefficient $\gamma^{calc}$ of 5.8 mJ/mol K$^2$ in LaRu$_2$P$_2$ and 10.8 mJ/mol K$^2$ in LaFe$_2$P$_2$. In LaRu$_2$P$_2$ we have measured a $\gamma^{exp}$ of 11.5 mJ/mol K$^2$, giving a mass enhancement factor $\lambda^\gamma$ defined as $\gamma^{exp}/\gamma^{calc} = (1+\lambda^\gamma)$ of 0.98. Similarly in LaFe$_2$P$_2$ we find a $\gamma^{exp}$ of 19.3 mJ/mol K$^2$ and $\lambda^\gamma$ of 0.8.

\begin{table*}[t]

\begin{tabular}{l || c | c || c | c | c}
Material & Measured cyclotron mass  & Renormalization $\lambda^{cyc}$ & Measured Specific Heat $\gamma$  & Calculated $\gamma$ & Renormalization $\lambda^\gamma$ \\
&(m/m$_{e}$) & from cyc. mass & (mJ/mol K$^2$) & (mJ/mol K$^2$) & from $\gamma$\\
\hline
\hline
LaRu$_2$P$_2$ & 0.71 - 0.99 & 0.8 & 11.5 & 5.8 & 1 \\
LaFe$_2$P$_2$ & 2.0-2.7 \cite{Muranaka} &   & 19.3  & 10.8 & 0.8 \\
\hline
LaRuPO & 0.55 - 0.86 \cite{Muranaka}&   & 3.9 \cite{Krellner} & 2 & 0.95 \\
LaFePO & 1.8-2.1 \cite{Coldea2} & 1 \cite{Coldea2} & 12.5 \cite{McQueen} & 6 & 1.08 \\
\hline
BaFe$_2$As$_2$ \cite{Analytis2} & 0.6-1.2 & 0.7-1 & 6.1 & 2.83 & 1.1 \\
SrFe$_2$As$_2$ \cite{Suchitra} & 1.5-2.0 &   & 3.3 & 3.5 & 0.85 \\
CaFe$_2$P$_2$ & 2.05 - 4.0 \cite{Coldea} & 0.45 - 0.51 \cite{Coldea} & 6.5 \cite{Chen} &   &  \\
BaNi$_2$P$_2$ \cite{Terashima} & 0.53 - 1.50 & 0.3 - 2.1 &   & 8.78 &  \\

\end{tabular}
\caption{Mass enhancement from cyclotron mass and specific heat in Fe and Ru pnictides.}
\label{Table1}
\end{table*}

\begin{figure}[h]
\centering
\includegraphics[width =0.45 \textwidth]{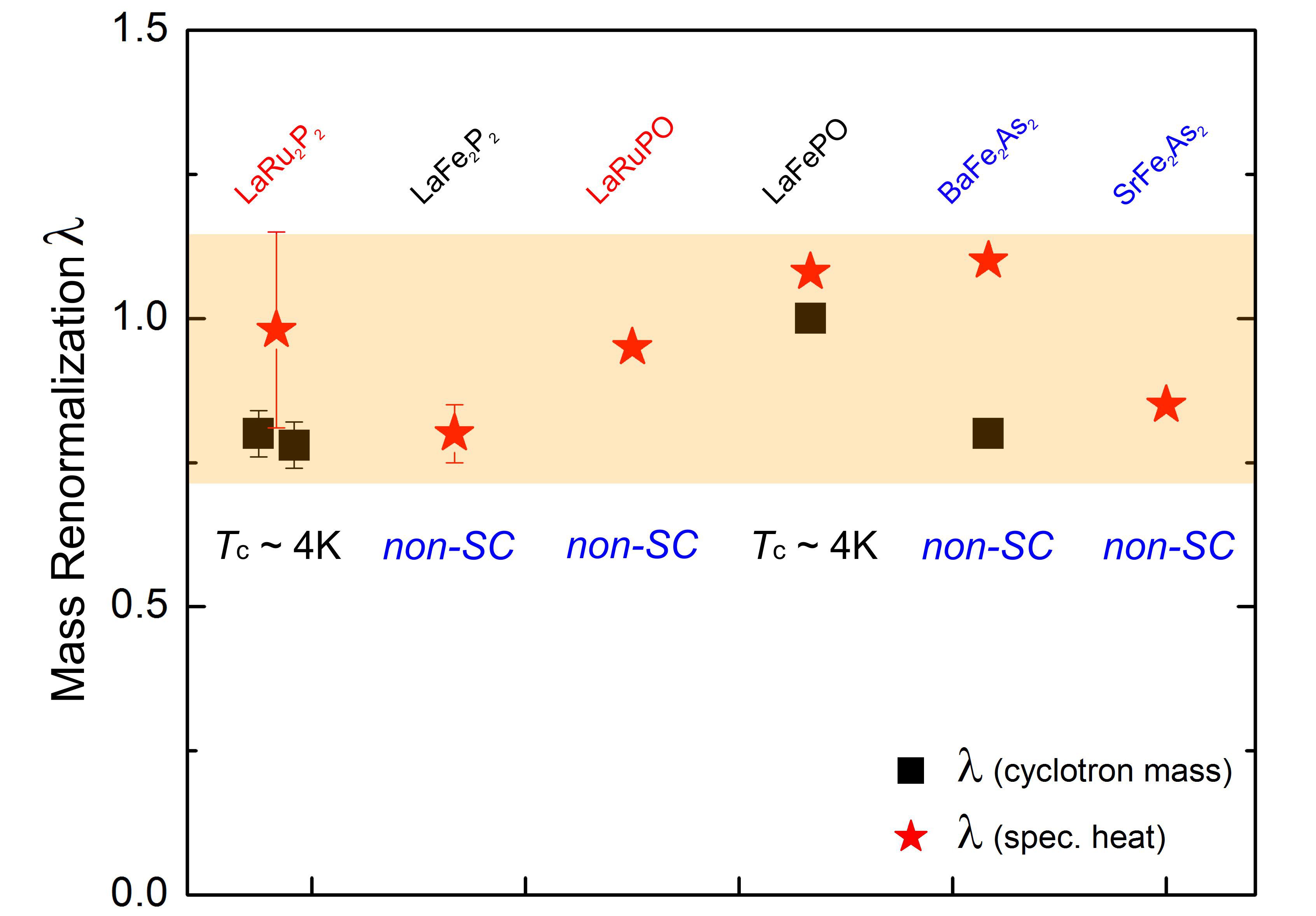}
\caption{Mass renormalization $\lambda$ of LaRu$_2$P$_2$ and other related compounds, determined by quantum oscillations and specific heat measurements. For all these materials $\lambda$ is similar around 1.}
\label{fig:Lambda}
\end{figure}

This should be compared to the effective mass enhancement of orbits on individual $m_{exp}/m_{b} = (1+\lambda^{cyc})$. We have extracted the band cyclotron mass $m_{b}$ for each orbit $S_k$ and field angle from our calculated band structure using the relation $m_{b} = \hbar /2\pi e$ $dS_{k}/dE|_{E_F}$. At $\theta$ = 20\textdegree, we find $m_{b}^\alpha$ = 0.40 $m_e$ and $m_{b}^\beta$ = 0.55 $m_e$ leading to a mass enhancement $\lambda^{cyc}$ of 0.78 for $\alpha$ and 0.8 for $\beta$ . If the small difference between $\lambda^{cyc}$ and $\lambda^{\gamma}$ is taken literally, it suggests that the unobserved three dimensional sheet is more renormalized compared to the donut and cylinder sheets.\\
Overall, this yields a highly consistent picture of similar mass enhancements with $\lambda \approx 1$ in the iron-pnictide compounds upon Ru substitution, even as LaRu$_2$P$_2$ and LaFePO are superconductors and LaFe$_2$P$_2$ and LaRuPO are not (sketched in Fig. \ref{fig:Lambda} and Table \ref{Table1}). The substitution of Fe for Ru was recently studied by ARPES measurements \cite{Brouet} and it was suggested to reduce the electronic correlations in the '122' pnictides. It is impossible in dHvA experiments to separate the electron-electron interaction enhancement from the electron-phonon contribution. However, because the total mass renormalization is similar in the Ru and the Fe compounds, it seems to be a remarkable coincidence that the reduction of electronic correlations due to the broader Ru bands would be compensated by an equivalent increase in electron-phonon interactions. It will be of interest to compare these mass renormalizations at E$_F$ with the optically probed mass enhancement at higher energies \cite{Quazilbash}.\\

In conclusion, our dHvA studies and specific heat measurements reveal that LDA calculations reproduce the band structure of LaRu$_2$P$_2$ in great detail and give highly similar Fermi surfaces for LaRu$_2$P$_2$ and LaFe$_2$P$_2$. Thus differences in FS nesting are unlikely the explanation for the appearance of superconductivity in LaRu$_2$P$_2$. A similar, sizeable quasiparticle mass renormalization $\lambda \approx 1$ is found in both discussed pairs LaRu$_2$P$_2$({\it T}$_c$ = 4K) --– LaFe$_2$P$_2$ and LaRuPO -- LaFePO({\it T}$_c$ = 4K), as well as other pnictides. This similar mass enhancement observed in both superconducting and non superconducting pnictides is contrary to the common perceptions that the normal state quasi-particle renormalizations reflect the strength of the superconducting paring mechanism.

\begin{acknowledgments}
This work was supported by the Swiss National Science Foundation and the National Center of Competence in Research MaNEP (Materials with Novel Electronic Properties). Work at NHMFL-LANL is carried out under the auspices of the National Science Foundation, Department of Energy and State of Florida. PB was supported by the Austrian Science Fund SFB-F41 (ViCoM).
\end{acknowledgments}

\bibliography{LaRu2P2-Bib}

\end{document}